\documentclass[sigplan,10pt]{acmart}
\renewcommand\footnotetextcopyrightpermission[1]{} % removes footnote with conference information in first column
\acmConference[arXiv]{}
\acmYear{2023}
\acmISBN{} % \acmISBN{978-x-xxxx-xxxx-x/YY/MM}
\acmDOI{} % \acmDOI{10.1145/nnnnnnn.nnnnnnn}
\startPage{1}

\setcopyright{none}
\usepackage{booktabs}   %% For formal tables:
\usepackage{subcaption} %% For complex figures with subfigures/subcaptions
\preto{\abstractkeywords}{\nolinenumbers} 

\usepackage{amsmath,amsfonts}
\usepackage{graphicx}
\usepackage{svg}
\usepackage{pdfpages}
\usepackage{textcomp}
\usepackage{stfloats}
\usepackage{xcolor}
\definecolor{mygray}{gray}{.8}
\usepackage{multirow}
\usepackage{colortbl}
\makeatletter
\newif\if@restonecol
\makeatother

\usepackage[ruled,vlined]{algorithm2e}%[ruled,vlined]{
\begin{document}

%% Title information
\title{Stencil Matrixization} %% [Short Title] is optional;
%% when present, will be used in
%% header instead of Full Title.
% \titlenote{with title note} %% \titlenote is optional;
% %% can be repeated if necessary;
% %% contents suppressed with 'anonymous'
% \subtitle{Subtitle} %% \subtitle is optional
% \subtitlenote{with subtitle note}   %% \subtitlenote is optional;
%% can be repeated if necessary;
%% contents suppressed with 'anonymous'

%% Author information
%% Contents and number of authors suppressed with 'anonymous'.
%% Each author should be introduced by \author, followed by
%% \authornote (optional), \orcid (optional), \affiliation, and
%% \email.
%% An author may have multiple affiliations and/or emails; repeat the
%% appropriate command.
%% Many elements are not rendered, but should be provided for metadata
%% extraction tools.
%% Author with single affiliation.
\author{Wenxuan Zhao}
% \authornote{with author1 note}  %% \authornote is optional;
%% can be repeated if necessary
\orcid{nnnn-nnnn-nnnn-nnnn} %% \orcid is optional
\affiliation{
  % \position{Position1}
  \department{Institute of Computing Technology,}  %% \department is recommended
  \country{} 
  \institution{Chinese Academy of Sciences}
  \institution{University of Chinese Academy of Sciences}%% \institution is required
   %% \country is recommended
}
\email{zhaowenxuan21@mails.ucas.ac.cn}  %% \email is recommended

%% Author with two affiliations and emails.
\author{Liang Yuan}
% \authornote{Corresponding author}  %% \authornote is optional;
%% can be repeated if necessary
\orcid{nnnn-nnnn-nnnn-nnnn} %% \orcid is optional
\affiliation{
  \department{Institute of Computing Technology,}  %% \department is recommended
  \institution{Chinese Academy of Sciences}
  \country{} 
}
\email{yuanliang@ict.ac.cn} %% \email is recommended

\author{Baicheng Yan}
  %% \authornote is optional;
%% can be repeated if necessary
\orcid{nnnn-nnnn-nnnn-nnnn} %% \orcid is optional
\affiliation{
  % \department{Institute of Computing Technology,}  %% \department is recommended
  \institution{Huawei Technologies Co., Ltd}
  \country{} 
}
\email{yanbaicheng@huawei.com} %% \email is recommended

\author{Penghao Ma}
  %% \authornote is optional;
%% can be repeated if necessary
\orcid{nnnn-nnnn-nnnn-nnnn} %% \orcid is optional
\affiliation{
  % \department{Institute of Computing Technology,}  %% \department is recommended
  \institution{Huawei Technologies Co., Ltd}
  \country{} 
}
\email{mapenghao@hisilicon.com} %% \email is recommended
\author{Yunquan Zhang}
  %% \authornote is optional;
%% can be repeated if necessary
\orcid{nnnn-nnnn-nnnn-nnnn} %% \orcid is optional
\affiliation{
  \department{Institute of Computing Technology,}  %% \department is recommended
  \institution{Chinese Academy of Sciences}
  \country{} 
}
\email{zyq@ict.ac.cn} %% \email is recommended

\author{Long Wang}  %% \authornote is optional;
%% can be repeated if necessary
\orcid{nnnn-nnnn-nnnn-nnnn} %% \orcid is optional
\affiliation{
  % \department{Institute of Computing Technology,}  %% \department is recommended
  \institution{Huawei Technologies Co., Ltd}
  \country{} 
}
\email{wanglong82@huawei.com} %% \email is recommended

\author{Zhe Wang}
  %% \authornote is optional;
%% can be repeated if necessary
\orcid{nnnn-nnnn-nnnn-nnnn} %% \orcid is optional
\affiliation{
  % \department{Institute of Computing Technology,}  %% \department is recommended
  \institution{Huawei Technologies Co., Ltd}
  \country{} 
}
\email{wangzhe125@huawei.com} %% \email is recommended
%% Abstract
%% Note: \begin{abstract}...\end{abstract} environment must come
%% before \maketitle command
\begin{abstract}

Current architectures are now equipped with matrix computation units designed to enhance AI and high-performance computing applications. Within these architectures, two fundamental instruction types are matrix multiplication and vector outer product, with the latter being lighter due to its vector inputs. This characteristic not only allows for the development of flexible algorithms beyond dense linear algebra computations but also offers greater potential for implementation optimization.

Stencil computations, commonly found in scientific and engineering applications, involve nested loops. This paper introduces a novel stencil algorithm leveraging vector outer products. Unlike previous approaches, this algorithm emerges from the stencil definition in scatter mode and is initially formulated using vector outer product expressions. The implementation integrates a series of optimizations to enhance memory reference patterns, execution pipeline efficiency, and data reuse. These optimizations consider various algorithmic options and data sharing among input vectors.

Evaluation conducted on a simulator demonstrates that our proposed design achieves significant speedup compared to vectorized stencil algorithms.
\end{abstract}

%% 2012 ACM Computing Classification System (CSS) concepts
%% Generate at 'http://dl.acm.org/ccs/ccs.cfm'.

%% End of generated code

%% Keywords
%% comma separated list
  %% \keywords are mandatory in final camera-ready submission

%% \maketitle
%% Note: \maketitle command must come after title commands, author
%% commands, abstract environment, Computing Classification System
%% environment and commands, and keywords command.
\maketitle

%%%%%%%%%%%%%%%%%%%%
%%%%%%%%%%%%%%%%%%%%
%%%%%%%%%%%%%%%%%%%%
%%%%%%%%%%%%%%%%%%%%

%%%%%%%%%%%%%%%%%%%%
%%%%%%%%%%%%%%%%%%%%
%%%%%%%%%%%%%%%%%%%%
%%%%%%%%%%%%%%%%%%%%
\section{Introduction}

Matrix registers and associated operations have recently been equipped in processors as powerful units for accelerating modern AI and high-performance computing applications. 
As Matrix Multiplication (GEMM) is a fundamental computation pattern in  these applications,
many commercial hardware accelerators,
such as NVIDIA Tensor Cores~\cite{nv.tensorcore}, Google TPUs~\cite{google.tpu} and Intel AMXs~\cite{Intel.AMX},
have been designed 
to incorporate matrix-matrix multiplication instructions to offer $O(n^3)$ arithmetic operations with $O(n^2)$ operands.
The arithmetic density,
i.e, the number of floating point operations performed per byte is $O(n)$, 
one order of magnitude greater than
other traditional instructions.

% To meet the increasing demands of high throughput GEMM, 

% blas 2
% vector outer products
% blas 3
% gemm

It is well-known that GEMM can be implemented with inner products
or outer products. 
Inner product instructions exist in some architectures, e.g., ARMv8.4.
However, it only completes  $O(n)$ arithmetic operations.
On the contrary, the outer product
performs $O(n^2)$ calculations in a single instruction.
It can also be used as the building block of other computations such as triangular solvers.
Several architectures have proposed 
outer product accelerator units in their processors,
such as the Scalable Matrix Extension (SME) in ARM~\cite{ARM.SME}
and the Math Matrix Accelerator (MMA) in IBM Power10~\cite{IBM.MMA}.

 % Usually gemm is reducible to the outer product
 % i.e, any algorithm implmented with gemm
 % can directly utilize the outer products unit
 % by replacing gemm with a series of outer product
 % operations.
 
% However,  however, gemm input is matrix, not flexible as vector,
%  so with more fine-grained parallelism, it offers
%  more design opportunities.
%  outer product
% instructions are more fine-grained than a complete matrix multiply unit and .
% The fexiblity also reflect in the data organization.
% the vector reorganization is feasible than matrix.
% outer product
% can reuse the vector register file
% as input vector.

The outer product is fed with two vectors and produces a matrix.
Each element in the output matrix is a product of corresponding values in the two input vectors. %, as shown in formula 1 and Figure 1 . 
The outer product is clearly more lightweight compared with the matrix multiplication. 
It provides more opportunities to develop flexible algorithms for more problems other than dense linear algebra computing,
and more possibilities to optimize the instruction pipelines,  data organization
and data reuses among input vectors.
% A detailed discussion is present in Section \ref{sec-observation}.

% This set of new instructions includes data movement from SVE to SME, data movement from SME to SVE and memory, vector outer product calculation, etc. The vector is stored in SVE. The matrix is in SME. 

% Multiplication and addition can be realized simultaneously in SME by multiple sets of vector outer product operations, which makes it possible to use SME to optimize stencil calculations.

The stencil computation is identified as one of the thirteen Berkeley
motifs~\cite{Asanovic+:TR06} and represents a very common class of nested loops in scientific and engineering applications, dynamic programming, and
image processing algorithms.
% Each stencil has a fixed calculation mode, and there are many differences in the calculation modes of different stencils.
Stencils can be classified in terms of the dimension of the space grid (1D, 2D...),  the shape (box, star...)
and the order.
% In addition to some stencils of regular shape and dimension, there are also stencils that are irregular.

Existing work on stencil computation mainly focuses on  tiling~\cite{Bandishti+:sc12,Tang+:spaa11,Yuan+:sc17}, parallelization~\cite{Krishnamoorthy+:pldi07,Datta+:sc08,Zhang.Mueller:cgo12} and vectorization~\cite{Henretty+:cc11,Yount:hpcc15,yuan2021temporal}. Several pieces of literature discuss the stencil algorithms on GPU~\cite{Grosser+:cgo14,Nguyen+:sc10,Lutz+:taco13} and other
many-core processors~\cite{Cui+:jcst10},
and the transformation of stencil
computation to  matrix multiplications~\cite{Liu+:ics22,Moreira+:arxiv21}.
To the best of our knowledge, 
there is no stencil algorithm targeting the vector
outer product accelerator.
% The main purpose of optimization is to reduce redundant calculations, improve data reusability, optimize memory access, and improve instruction pipeline.

Most of the existing work adopts the conventional
gather view of a stencil definition,
where each output element is calculated
 by summing its neighbor input elements.
This perspective makes it difficult
to utilize outer products for stencil computations.
The scatter view of a stencil~\cite{Zhao+:sc19} %exploiting reuse and vectorization
describes the updating pattern of one single input element.
This standpoint has been used for
 fine-grained in-core optimizations,
 e.g., improving the data reuse~\cite{Cruz.Araya-polo:toms14} and
eliminating the register spilling~\cite{Stock+:pldi14}
by combining the gather and scatter views
for a single dimension and a single element, respectively.

% In addition, the load vector method is more flexible for memory reading, and a more memory-friendly algorithm can be designed to make full use of hardware prefetch to increase cache locality as much as possible.

This paper proposes a novel stencil algorithm using vector outer products.
Unlike previous work, the new algorithm 
 arises from the stencil definition in the scatter mode.
The algorithm for a 2D stencil is initially expressed with formulas of vector outer products by a series of extensions to the scatter mode.
Then it is adapted to other stencil types by identifying a key concept underlying the algorithm, and
it is also analyzed
theoretically concerning this concept.
The implementation  incorporates  
a set of optimizations
to improve the memory reference pattern, execution pipeline and data reuse
by considering various algorithmic options and the data sharing
between input vectors.
This paper makes the following contributions:

% Another advantage is the
% data reuse.
% Furthermore,
% for box stencil and $2$-dimensional stencils,
% the single formula target one subblock the output matrix
% allow the implementation
% to only write the output matrix to memory once.
% This resembles the strength of algorithms with the gather mode.

% Thus one input vector of $A$ is only scattered to one output vector.
% Our formula arises from the scatter mode
% that makes a full utilization of each input element.

 % the memory reference and execution pattern

\begin{itemize}
\item We propose a stencil algorithm using vector outer products.
The theoretical aspects of the algorithm
are also analyzed.

\item We implement the algorithm with a set of optimizations to boost the in-core performance. We also design
 an automatic code generator.

\item We demonstrate the effectiveness of the proposed algorithm, showing substantial performance improvement over a range of 2D and 3D stencils.
\end{itemize}

The remainder of this paper is organized as follows. Section 2 provides some background on matrix accelerator units
and stencil computations. Section 3 describes the stencil algorithm 
 that utilizes vector outer product operations,
 and also offers a detailed theoretical analysis.
 Section 4 presents a set of optimizations on
 the memory reference and data reuse.
 Section 5  evaluates the performance on a simulator using various 2D and 3D stencils.
 Section 6 reviews related work and Section 7 summarizes the paper.

 % experimental results during the process, and the comparison between our experimental results and the benchmark. 

 %https://arxiv.org/pdf/2206.02874.pdf
\section{Background}

\subsection{Vector and Matrix Extension}

 The data-level parallelism plays a vital role in boosting the performance of a single CPU.
 Hardware developments have produced a large number of various
  vector extensions to common CPU architectures,
  e.g., SSE, AVX, AVX2 and AVX-512 to X86,
   NEON and SVE to ARM,
   RVV to RISC-V.

To further meet the increasing demands of high throughput GEMM, many commercial CPU and accelerators have incorporated 
special extensions to perform matrix computations.
Generally there are two architecture types.
The first one provides
matrix-matrix multiplication units
supported for various matrix sizes.
Intel's Advanced Matrix Extensions~\cite{Intel.AMX},
Nvidia's Tensor Cores~\cite{nv.tensorcore}
and Google's Tensor Processing Units~\cite{google.tpu}
are  representatives.

The other type incorporates lightweight
vector outer product unit.
IBM's Math Matrix Accelerator (MMA)~\cite{IBM.MMA},
Apple's Matrix Coprocessor
and ARM's Scalable Matrix Extension (SME)~\cite{ARM.SME}
fall into this type.
SME offers a maximal vector length of 1024-bit,
longer than the 128-bit vector in MMA.
Futhermore, each matrix register in
MMA occupies the storage of 
eight vector registers.
Thus SME allows more optimization opportunities
with more vector and matrix registers.
% For example, it is desirable to allocate more
%  registers
% as input vectors while keeping additional
% storage for large output matrices.
In this work we only focus on the SME
 of  ARM.

\subsection{Stencil Computation}

A stencil is defined on a structured physical grid,
therefore the variables on grid points are generally coded with arrays.
Moreover, stencil computation
involves a temporal evolution of the variable
associated with each grid point.
Specifically, it computes the value at time $t+1$
using neighbor values at time $t$.
Usually, the implementation utilizes two array copies $A$ and $B$.
The stencil computation along the time dimension
reuses the two data copies as the input and output array alternatively.
This work only considers the algorithm inside one-time step
from time $t$ to $t+1$
and always uses $B$ as the output array at time $t+1$  
and $A$ the input array at time $t$.

The vectorization of stencil computation is straightforward
since the data dependences are all carried by
the outmost time loop.
Generally, one can rely on compilers to
generate codes that utilize hardware SIMD units.
The matrixization of stencil computation is also studied recently \cite{Liu+:ics22,Moreira+:arxiv21}.
Both are from the conventional gather view
of stencil.

\section{Design}

\subsection{Observations}
\label{sec-observation}

We  present several 
observations on the current architecture specification,
including the matrix storage,
matrix accumulation and vector product.
They provide insights into the algorithm design in this section and algorithm implementation described in the next section.

Firstly, the matrix register assembling is achieved
either by the vector register to matrix register movement or
the memory load/store.
Both approaches operate at the vector granularity,
i.e., they demand $n$ instructions to fully
fill all the $n^2$ elements in a matrix register.
There are no available intra- or inter-matrix register re-organization instructions. 
For example, to 
transpose a matrix register on current
architectures, it has to 
move all its columns to vector registers (or memory) and
then move them back to rows of another matrix register.
Thus it  either consumes more vector registers or incurs
more memory references.
On the other side, the architectures often
provide substantial data re-organization instructions for vectors.
Consequently assembling data with vector registers is cheap and flexible.

Secondly,
the outer product instruction
 involves a matrix register
 as both input and output.
Each element in the matrix register
is added with the corresponding
 result from the outer product.
Thus the outer product
 offers a capability of $n^2$ addition operations
 besides the $n^2$ multiplications.
For stencil kernels targeted in this work, the numbers of multiplication and addition operations are approximately equal.
Therefore, it is desirable to  map all the stencil computations to  outer products
and fully utilize the $2n^2$ flops per instruction.

Thirdly,
 compared with the matrix-matrix multiplication instruction,
 the outer product operation
only provides a theoretical $O(1)$ arithmetic density
due to the size of the output matrix.
However, if the output matrix 
serves as an intermediate accumulator matrix
that is reused by a set of outer products.
The outer product can be  still
viewed as a high
computation intensity, $O(n^2)$ operations
over input vectors of $O(n)$ size.

The above three observations 
motivate us to reduce the movement
of output elements and keep them in a matrix
register as long as possible.
Meanwhile,
we resort to employing the rich set of vector
re-organization operations to assemble
demanded vectors for updating the fixed output matrix.
% The matrix register
% should be maximal updated.

Finally, an outer product operation creates a 2-dimensional matrix from two one-dimensional vectors.
The operation follows a scatter mode
where one single element in an input vector is scattered to a column or a row in the output matrix. 
This motivates the reformulation of
the stencil computation
in a similar scatter mode.
It also indicates that 
when considering
 the outer product operation related to 
 physical grid,
 it is meaningful 
only if the two input vectors  
are linearly independent
in terms of their directions.
With this view,
our algorithm is not 
applicable to one-dimensional stencils.

\subsection{Basic Formula}

A stencil represents an updating pattern of neighbor points.
It is often defined in a $gather$ mode,
i.e., a formula identifying the calculation of a new point $B_{i,j}$
using its neighbor points from  $A$. 
Equation~(\ref{stencil-2d9p})
shows the 2D9P stencil.
The new value at point $(i,j)$ on the two-dimensional grid
is calculated by multiplying its 9 neighbors with corresponding coefficients
and gathering the results with a reduction operation.

\begin{equation}
\label{stencil-2d9p}
\resizebox{0.9\hsize}{!}{
$\begin{aligned}
B_{i,j} &= C_{00} \times A_{i-1,j-1}&+ C_{01} \times A_{i-1,j}&+ C_{02} \times A_{i-1,j+1} \\
&+ C_{10} \times A_{i,j-1\phantom{+1}}&+C_{11} \times A_{i,j\phantom{+1}} &+ C_{12} \times A_{i,j+1\phantom{+1}} \\
&+ C_{20} \times A_{i+1,j-1}&+ C_{21} \times A_{i+1,j}&+ C_{22} \times A_{i+1,j+1}
\end{aligned}$}
\end{equation}

Consequently, the 2D9P stencil
can be simply identified by its coefficient matrix
in gather mode $C^g$, as shown in Equation~(\ref{stencil-2d9p-cm}).
$C^g$ is a $(2r+1)\times (2r+1)$ matrix, where $r$ is the stencil order.
For the 2D9P stencil, we have $r=1$.

\begin{equation}
\label{stencil-2d9p-cm}
\resizebox{0.45\hsize}{!}{
 $C^g$ =$ \left(
   \begin{array}{ccc}
  {C}_{00} & {C}_{01} & {C}_{02} \\
  {C}_{10} & {C}_{11} & {C}_{12} \\
  {C}_{20} & {C}_{21} & {C}_{22}
   \end{array}
   \right) $}
\end{equation}

In essence, the gather representation resembles an inner product of 
two vectors. One vector contains all the 9 coefficients and
the other one assembles corresponding neighbors for each output grid point.
It opposites the principle of outer product of vectors.
First, it is impossible to assemble an output matrix of $B$ that gives rise to
 reasonable vector organizations for coefficients and input values.
Second, it cannot perform the addition operation at the input vector side with outer product instructions.
Since the outer product is the basic target operation of this work, 
it is hard to develop an algorithm from this perspective.

Fortunately, a stencil can be described in a $scatter$ mode.
The scatter mode shows how to use just one point in the input array
to update its neighbors in the output array.
Equation~(\ref{stencil-2d9p-scatter})
shows the 2D9P stencil in the scatter mode.
The input value $A_{i,j}$ is multiplied with coefficients and
the results are accumulated to its corresponding  neighbor points in the output array.
The symbol $\odot$ denotes an element-wise multiplication.

\begin{equation}
\label{stencil-2d9p-scatter}
\resizebox{0.9\hsize}{!}{
  $\left(
   \begin{array}{ccc}
  {B}_{i-1,j-1} & {B}_{i-1,j} & {B}_{i-1,j+1} \\
  {B}_{i,j-1\phantom{+1}} & {B}_{i,j\phantom{+1}} & {B}_{i,j+1\phantom{+1}}  \\
  {B}_{i+1,j-1} & {B}_{i+1,j} & {B}_{i+1,j+1} 
   \end{array}
   \right) \xleftarrow{accum}
 \left(
   \begin{array}{ccc}
  {C}_{22} & {C}_{21} & {C}_{20} \\
  {C}_{12} & {C}_{11} & {C}_{10} \\
  {C}_{02} & {C}_{01} & {C}_{00}
   \end{array}
   \right)$ $\odot$ $A_{i,j}$  }
\end{equation}

Similarly, the 2D9P stencil
can be simply identified by its coefficient matrix
in scatter mode $C^s$, as shown in Equation~(\ref{stencil-2d9p-sm}).

\begin{equation}
\label{stencil-2d9p-sm}
\resizebox{0.45\hsize}{!}{
 $C^s =   \left(\begin{array}{ccc}
  {C}_{22} & {C}_{21} & {C}_{20} \\
  {C}_{12} & {C}_{11} & {C}_{10} \\
  {C}_{02} & {C}_{01} & {C}_{00}
   \end{array}\right)$}
\end{equation}

The coefficient matrix
in scatter mode $C^s$
can be obtained by 
reversing the rows and columns of 
the coefficient matrix
in gather mode $C^g$,
as shown in the following Equation.

\begin{equation}
\label{stencil-2d9p-gm2sm}
\resizebox{0.4\hsize}{!}{
 $C^s$ 
   =$J_{2r+1} \times C^g\times J_{2r+1}$}
\end{equation}

Here $J_{2r+1}$ is the \emph{reversal matrix},
where the 1 elements reside on the anti-diagonal and all other elements are zero.
It is a special case of permutation matrix
and $J_3$ for the 2D9P stencil is shown as follows.

\begin{equation}
\label{stencil-2d9p-jmatrix}
\resizebox{0.3\hsize}{!}{
 $J_3 
   =\left(
   \begin{array}{ccc}
  0 & 0 & 1 \\
  0& 1 & 0 \\
  1& 0 & 0 
   \end{array}
   \right)$}
\end{equation}

Since the outer product takes vectors as input,
we limit the scope to just one column of the output matrix and coefficient matrix in the scatter mode.
The following formula picks the middle column of $B$ and $C^s$ of Equation~(\ref{stencil-2d9p-scatter}).

\begin{equation}
\label{stencil-2d9p-scatter-2}
\resizebox{0.55\hsize}{!}{
  $\left(
   \begin{array}{ccc}
  {B}_{i-1,j}  \\
   {B}_{i,j\phantom{+1}}   \\
   {B}_{i+1,j}
   \end{array}
   \right) \xleftarrow{accum}
 \left(
   \begin{array}{ccc}
   {C}_{21}  \\
   {C}_{11}  \\
  {C}_{01} 
   \end{array}
   \right) \odot$ $A_{i,j}$  }
\end{equation}

The final formula that completes the computation of a $n\times n$ subblock of $B$ for a 2-dimensional stencil with order $r$ is developed
from Equation~(\ref{stencil-2d9p-scatter-2}) 
by three extensions:
 horizontally stretching the output value from a
$3\times 1$ vector to a $3\times n$ matrix,
gathering all the contributions
to transfer the accumulation arrow to an equal symbol,
and  vertically stretching the involved
$3\times 1$ vectors to  $n\times 1$ vectors with a parameterized stencil order $r$.
Combining these extensions, we obtain the final formula that
can be directly processed with a series of vector outer product operations.

Firstly, enlarging Equation~(\ref{stencil-2d9p-scatter-2})  horizontally is identical to extending the vector-scalar product $\odot$ to a vector outer product $\otimes$. It is achieved by simply vectorizing
each element of the $B$ vector and $A_{i,j}$ 
horizontally to $n$ elements
along the $j$ dimension, as shown in  Equation~(\ref{to-outer-product}).
We can also derive this formula by grouping
$n$ contiguous updates of Equation~(\ref{stencil-2d9p-scatter-2}).
This formula can be directly processed by 
a single vector-vector outer product instruction.

\begin{equation}
\label{to-outer-product}
\resizebox{0.9\hsize}{!}{
 $ \left(
   \begin{array}{ccc}
   {B}_{i-1,j} & \dots &{B}_{i-1,j+n-1}  \\
   {B}_{i,j\phantom{+1}} & \dots &{B}_{i,j+n-1\phantom{+1}}   \\
   {B}_{i+1,j} & \dots &{B}_{i+1,j+n-1}   
   \end{array}
   \right) \xleftarrow{accum}
 \left(
   \begin{array}{ccc}
  {C}_{21} \\
   {C}_{11}  \\
   {C}_{01} 
   \end{array}
   \right) \otimes$ $(A_{i,j},\dots,A_{i,j+n-1})$  }
\end{equation}

Secondly, 
we seek to extend the accumulation arrow in Equation
(\ref{stencil-2d9p-scatter-2}) to an equal symbol.
According to the stencil
definition in the scatter mode,
there are  other
4 input elements,
$A_{i-2,j}$,
$A_{i-1,j}$,
$A_{i+1,j}$ and 
$A_{i+2,j}$
that should be scattered to 
the output vector
$({B}_{i-1,j},{B}_{i,j},{B}_{i+1,j})^T$
with corresponding coefficient
vectors containing $C_{01}$, $C_{11}$ and $C_{21}$.
The following formula
summarizes all these contributions
and we refer to this summation as the \emph{coefficient line summation} $CLS(*,1)$, where $(*,1)$ indicates all the coefficients in the second column of $C^s$.
The accumulation arrow in Equation
(\ref{stencil-2d9p-scatter-2}) can be 
 replaced with an equal symbol by summarizing over all coefficient lines, i.e., $CLS(*,0)+CLS(*,1)+CLS(*,2)$.

\begin{equation}
\label{stencil-2d9p-scatter-extend-2}
\resizebox{0.9\hsize}{!}{
 $ \begin{aligned}
CLS(*,1) &= \left(
   \begin{array}{ccc}
  {C}_{01} \\
  0 \\
  0
   \end{array}
   \right)
   \times A_{i-2,j}
   +
   \left(
   \begin{array}{ccc}
  {C}_{11} \\
  {C}_{01} \\
  0
   \end{array}
   \right)
   \times A_{i-1,j}\\
   & +
   \left(
   \begin{array}{ccc}
  {C}_{21} \\
  {C}_{11} \\
  {C}_{01}
   \end{array}
   \right)
   \times A_{i,j} + \left(
   \begin{array}{ccc}
  0 \\
  {C}_{21} \\
  {C}_{11} 
   \end{array}
   \right)
   \times A_{i+1,j}
   +
   \left(
   \begin{array}{ccc}
  0\\
  0\\
  {C}_{21} 
   \end{array}
   \right)
   \times A_{i+2,j} 
\end{aligned}$
}
\end{equation}

Finally, we adapt Equation~(\ref{stencil-2d9p-scatter-2})
to a vector of size $n$ 
by filling zeros to the coefficient vector
and successive elements below $B_{i,j}$
to the output vector.
The reason is that 
in the 2D9P stencil,
each input element of A is at most scattered to 3 output elements with one coefficient line.  
In addition, we adapt Equation~(\ref{stencil-2d9p-scatter-2}) to 
a 2-dimensional box stencil with parameterized order $r$. It
matches a $(2r+1)\times(2r+1)$ coefficient matrix and consequently yields $2r+1$ coefficient lines with $2r+1$ weights in each line.
Similar to Equation~(\ref{stencil-2d9p-scatter-2}),
$A_{i+r,j}$ is scattered to $B_{i+k,j}$ with $C_{2r-k,1}$ for $0\leqslant k\leqslant 2r$.
The combination of the above two adaptations
leads to the following formula.

\begin{equation}
\label{stencil-2d9p-scatter-extend-3}
\resizebox{0.6\hsize}{!}{
 $   \left(
   \begin{array}{ccc}
  {B}_{i,j\phantom{+2r+1}}  \\
   \dots   \\
   {B}_{i+2r,j\phantom{+1}} \\
   \dots\\
   \dots\\
   {B}_{i+n-1,j\phantom{2}}
   \end{array}
   \right)\xleftarrow{accum}
 \left(
   \begin{array}{ccc}
   {C}_{2r,1}  \\
   \dots  \\
  {C}_{0,1} \\
  0\\
  \dots\\
  0
   \end{array}
   \right) \odot$ $A_{i+r,j}$  }
\end{equation}

To simplify the description
of the final formula,
we introduce the
outer product coefficient matrix 
$C^o$ as shown in Equation~(\ref{co}).
$C^o$ expands the coefficient matrix
in the scatter mode $C^s$
by adding two zero matrices of 
size $(n-1)\times(2r+1)$ below and above it.
For example, the second column of $C^o$
is the vector
$(0,\dots,0,C_{2r,1},\dots,C_{0,1},0,\dots,0)^T$.
According to Equation~(\ref{stencil-2d9p-scatter-extend-2}),
every coefficient vector that 
participates in the computation
is a sub-sequence of such a one-column vector.

\begin{equation}
\label{co}
\resizebox{0.5\hsize}{!}{
   $C^o =  \left(
   \begin{array}{ccc}
  \Large{0}_{n-1,2r+1} \\
  C^s_{2r+1,2r+1} \\
  0_{n-1,2r+1}
   \end{array}
   \right)_{2n+2r-1,2r+1} $ }
\end{equation}

Put it all together,
the final formula is shown in 
Equation~(\ref{final}).
The outer summation with
the index $j$  corresponds to
$2r+1$ columns of $C^o$,
or equivalently, coefficient lines
of $C^s$.
For each coefficient line,
the inner summation with the index $i$
gathers all $(2r+n)$ related  vectors form the input matrix $A$ and  matrix $C^o$.

\begin{equation}
\label{final}
\resizebox{0.8\hsize}{!}{
 $ \begin{aligned}
&  \left(
   \begin{array}{ccc}
   {B}_{r,r} & \dots &{B}_{r,r+n-1}  \\
& \dots &   \\
   {B}_{r+n-1,r} & \dots &{B}_{r+n-1,r+n-1}   
   \end{array}
   \right)= \\
& \sum_{j=0}^{2r}\sum_{i=0}^{n+2r-1} \left(
   \begin{array}{ccc}
  C^o_{n+2r-i-1,j} \\
\dots \\
  C^o_{2n+2r-i-2,j}
   \end{array}
   \right)\otimes (A_{i,j},\dots,A_{i,j+n-1})
   \end{aligned} $ }
\end{equation}

\subsection{Adapt to Various Stencils}

The essential concept
underlying  the basic formula shown above
is 
the coefficient line.
The rest of this section discusses
the adaption of it to various stencils
and presents a theoretical analysis.

\emph{High dimension stencils}. 
For 
higher-dimensional box stencils,
the coefficients in the scatter mode
form a tensor and they
can be grouped into a set of coefficient lines similarly.
Take the 3D27P stencil as an example,
where the coefficients are 
$C_{i,j,k}$ $(i,j,k=0,1,2)$,
it consists of 9 
coefficient lines
$CLS(i,*,k)$.
% and each one
% comprises  $C_{i,*,k}$.
Adapting Equation~(\ref{final})
to the 3D27P stencil
is straightforward.
The output matrix is 
$B_{i,r:(r+n-1),r:(r+n-1)}$
and the input vector of $A$
is $A_{i',j',k':(k'+n-1)}$
where $i-r\leqslant i'\leqslant i+r$,
$0\leqslant j'< n+2r$
and $0\leqslant k'\leqslant 2r$.
The detailed formula is similar and is thus omitted.

\emph{Star stencils}. A star stencil can be regarded as
 a box stencil
 by setting corresponding weights that are empty in the  star stencil
 to zero.
 The following formula presents
 the coefficient matrix in the scatter mode
 of the 2D5P star stencil.
 Therefore it can directly employ
 Equation~(\ref{final})
 to perform the stencil computation.

\begin{equation}
\label{stencil-2d5p-sm}
\resizebox{0.45\hsize}{!}{
 $C^s =   \left(\begin{array}{ccc}
  0 & {C}_{21} & 0 \\
  {C}_{12} & {C}_{11} & {C}_{10} \\
  0 & {C}_{01} & 0
   \end{array}\right) $ }
\end{equation}

However, the first and third 
coefficient lines only contain
one nonzero 
and consequently each vector outer production
in these coefficient  line summations
is degraded to a 
scalar-vector multiplication.
Thus it does not 
exploit  the $O(n^2)$ arithmetic ability
of vector outer productions
and incur extra  overhead for
 the coefficient vector organization.

The aforementioned three extensions and final formula
 target the middle column of the coefficient matrix.
However, the concept of coefficient line 
is not restricted to a specific dimension.
It is legal to extract the second row
of $C^s$ in Equation~(\ref{stencil-2d5p-sm})
and those extensions can also be
applied to $(B_{i,j-1},B_{i,j},B_{i,j+1})$$=$$ A_{i,j}\odot (C_{12},C_{11},C_{10})$
similarly.
Note that the coefficient line lies
on the $j$ dimension, thus
the vector extension of $A_{i,j}$
should be along the $i$ dimension.
Equation~(\ref{to-outer-product-2-star})
provides the first extension. 
% Note that since the coefficient line
% is horizontal, the input vector
% is extended along the $i$
% dimension.
The rest deductions are similar
and thus the final formula is also omitted.

\begin{equation}
\label{to-outer-product-2-star}\resizebox{0.85\hsize}{!}{
  $  \left(
   \begin{array}{ccc}
   {B}_{i,j-1} & {B}_{i,j} &{B}_{i,j+1}  \\
   \dots & \dots & \dots  \\
   {B}_{i+n-1,j-1} & {B}_{i+n-1,j} &{B}_{i+n-1,j+1} 
   \end{array}
   \right) \xleftarrow{accum}
 \left(
   \begin{array}{ccc}
  A_{i,j} \\
   \dots  \\
   A_{i+n-1,j}  
   \end{array}
   \right) \otimes$ $({C}_{12},{C}_{11},{C}_{10})$  }
\end{equation}

The final formula for 2D star stencils 
can be obtained by merging the two accumulations,
 i.e., $CLS(*,1)+CLS(1,*)$.
 The final formula can be expressed with a single equation.
However, for $3$
and higher-dimensional star stencils,
accumulations of all coefficient lines
cannot be merged into a single formula.
% $CLS(*,1,1)$, $CLS(1,*,1)$, $CLS(1,1,*)$.
A detailed discussion will be provided in Section~\ref{sec-dataaccesspattern}.

\emph{Other Stencils}. Although this work mainly discusses 
common stencils like the star or box shape,
the concept of coefficient line is flexible
to handle other irregular shapes.
The following coefficient matrix 
has non-zero weights only on the main diagonal and anti-diagonal.
Thus it only requires two coefficient line summations, $CLS(diagonal)$ and
$CLS(antidiagonal)$.

\begin{equation}
\label{stencil-2d3p-sm}
\resizebox{0.45\hsize}{!}{
 $C^s =   \left(\begin{array}{ccc}
  {C}_{22} & 0 & {C}_{20}  \\
  0 & {C}_{11} & 0 \\
   {C}_{02} & 0 & {C}_{00}
   \end{array}\right) $ }
\end{equation}

For the anti-diagonal, the initial equation in the scatter mode is
shown in Equation~(\ref{stencil-2d3p-scatter-2}).
Note that the elements in the output vector,
i.e., ${B}_{i-1,j+1}$, ${B}_{i,j}$
and ${B}_{i+1,j-1}$ in the output $B$ matrix form the same direction 
to the anti-diagonal.
Similarly, the extensions are also applicable to 
Equation~(\ref{stencil-2d3p-scatter-2})
and the final formula is omitted.

\begin{equation}
\label{stencil-2d3p-scatter-2}
\resizebox{0.55\hsize}{!}{
  $\left(
   \begin{array}{ccc}
  {B}_{i-1,j+1}  \\
   {B}_{i,j\phantom{+1}}   \\
   {B}_{i+1,j-1}
   \end{array}
   \right) \xleftarrow{accum}
 \left(
   \begin{array}{ccc}
   {C}_{20}  \\
   {C}_{11}  \\
  {C}_{02} 
   \end{array}
   \right) \odot$ $A_{i,j}$  }
\end{equation}

\subsection{Analysis}
One outer production performs
$O(n^2)$ multiply-and-add operations.
Thus we expect a $1/n$ decrease
in terms of the instruction number.
Since the vectorization unit also provides
the fused multiply-and-add instructions,
it is reasonable to consider only the 
multiplication operations.
For 2D box stencils,
the total number of multiplications is $ N^2\times(2r+1)^2$,
where $N^2$ is the problem size.
With vectorization, each instruction
performs $n$ multiplications simultaneously.
The vector instruction number is then divided by $n$,
i.e., $N^2\times(2r+1)^2/n$.
According to Equation~(\ref{final}),
each subblock $B_{n\times n}$
incur $(2r+1)\times(2r+n)$
 outer product operations
and there are total $N^2/n^2$ subblocks of
the output matrix.
Thus, the average instruction number per output vector 
decreases 
from $2r+1$
to $2r/n+1$.

This decrease can also be derived from 
the perspective of the coefficient line.
Each line contains $2r+1$ weights.
Using the outer product,
Equation~(\ref{final}) shows that
one line leads to $2r+n$ coefficient vectors
for processing $n$ row vectors in the output matrix $B$.
Therefore, the instruction related to each weight 
is reduced to $(2r + n)/n$.
This interpretation is also applicable to 
higher-dimensional box stencils.

For $d$-dimensional star stencils,
the number of nonzero weights is 
$(2r\cdot d+1)$, while the number is $(2r+1)^d$
for box stencils.
Each coefficient line for the star stencil
still incur  $2r+n$  vector productions
and there are total $d$ lines.
The average number of outer production
is $d(2r+n)/n$ per output vector,
compared with $2r\cdot d+1$
vectorization instructions.
Thus the decrease is from
$2r+1/d$ to $2r/n+1$.
This decrease is smaller than box stencils
and the reason is that the middle weight, e.g.,
$C_{11}$ in $C^s$ (Equation~(\ref{stencil-2d5p-sm})) for
the 2D5P star stencil,
only exists in one coefficient line.

\subsection{Minimal  Cover with Axis-parallel Coefficient Lines}

The final formula 
indicates that the optimal execution
in terms of the instruction number
desires a
minimal set of coefficient lines covering all the nonzero weights.
We neither find a 
polynomial algorithm
nor an NP-hard proof for this general minimal line cover problem.
% We leave it as an open problem.
However, restricting the scope to 2D stencils,
the minimal cover with axis-parallel coefficient lines is reducible to the minimum vertex cover problem of bipartite graphs.

% axis-parallel

% this work only considers regular stencils,
% therefore the line cover options are easy to obtain and
% analyze.

% However, there exist
% complex neighbor patterns 
% that require a general algorithm to .
% For 3D and higher-dimensional stencils,

In a bipartite graph $G$, the vertex set can be divided into two sets $U$ and $V$ such that $U$ and $V$ are disjoint, $U\cap V = \emptyset$, 
and complete, $U\cup V$ contains every vertex in $G$. An edge in the edge set $E$ of $G$ only connects one vertex from $U$ to another from $V$. Thus a bipartite graph can be denoted as $G(U,V,E)$.
A vertex cover of a graph is a set of vertices 
 such that every edge has at least one endpoint in the set.
The problem of finding a minimum vertex cover of a general graph is a well-known classical NP-hard optimization problem. However, for the bipartite graph, there exists a polynomial solution
based on the K\"{o}nig theorem
\cite{konigtheorem}.

% Consider a 2D stencil
% with order $r$,
% i.e., all nonzero coefficients
% belong to a matrix $C_{(2r+1)\times (2r+1)}$.
% According to Equation~(\ref{stencil-2d9p-gm2sm}),
% the minimal cover is identical to the coefficient matrices in
% both the gather and scatter modes.
% Similar to Equation~(\ref{stencil-2d9p}),
% we list the stencil in the gather mode
% as follows.

% \[B_{k,l} = \sum_{i=-r}^{r}\sum_{j=-r}^{r}C_{i+r,j+r} *A_{k+i,l+j}\]

Converting the minimal covering lines problem to the minimum vertex cover
of a bipartite graph is achieved 
by interpreting the coefficient matrix as the adjacency matrix of the bipartite graph $G(U,V,E)$. 
The vertex set $U$ contains a vertex $u_i$ if and only if there exists at least one $j\in\{0,\dots,2r\}$ such that $C_{i,j}$ is nonzero.
The other vertex set $V$ is constructed similarly.
Every nonzero coefficient $C_{i,j}$ is mapped to an edge $e_{i,j}$ 
that connects $u_i$ to $v_j$.
The minimum vertex cover $VC$ of $G(U,V,E)$ corresponds to 
a  minimal set of line cover $LV$.
Each $u_i\in VC$ implies a horizontal coefficient line in $LV$
and $v_i$  a vertical coefficient line.

\section{Implementation}

It is straightforward to 
write the stencil program
with the final formulas developed above.
However, the program's performance
is often bound by the data accesses.
This section discusses several
optimizations on memory references.
We rearrange the formula
to find a balance between 
the number of outer products and memory reference patterns,
and reschedule the operations of the formula
to obtain an efficient data reuse pattern
with a multi-dimensional unrolling.

% Conventional optimizations are also
% applicable, including loop unroll
% and data alignment.

\begin{table}[t]
\caption{Features of $CLS$ options for 2D star stencils}
\label{cls-2d}
\resizebox{\linewidth}{!}{
\begin{tabular}{cc|c|c}
\hline
\multicolumn{2}{c|}{Option}   & Input & \#Outer Product   \\ \hline
\multicolumn{1}{c|}{Parallel}& $CLS(*,j),j = 0,…,2r$ & $A_{1\times n}$ & $(2r+n)+2r\cdot n$   \\ \hline
\multicolumn{1}{c|}{\multirow{2}{*}{Orthogonal}} & $CLS(*,r)$& $A_{1\times n}$ & \multirow{2}{*}{$2(2r+n)$} \\ \cline{2-3}
\multicolumn{1}{c|}{}& $CLS(r,*$)& $A_{n\times 1}$ &  \\ \hline
\end{tabular}}
\end{table}

\subsection{Data Access Pattern}
\label{sec-dataaccesspattern}

Although an optimal coefficient line cover leads to the minimal outer product instruction number, the memory data reference
may be inferior. 
Take  the 2D5P 
star stencil as an example,
for calculating $CLS(*,1)$,
as shown in Equation~(\ref{to-outer-product}),
the elements in the input vector
$(A_{i,j},\dots,A_{i,j+n-1})$
are contiguous in memory.
\footnote{In this paper, we follow
a C-style storage.
We use $(i,j)$
for 2D stencils and  $(i,j,k)$ for 3D stencils, where
 the rightmost indices $j$ and $k$ refer to the contiguous accesses. 
%We also use $i$, $j$, $k$
%or $x$, $y$, $z$ to denote the dimension
%indices alternatively.
}
On the contrary, Equation~(\ref{to-outer-product-2-star})
indicates that
 the input vector
 in the 
$CLS(1,*)$ calculation
contains 
strided elements.
 Consequently $CLS(1,*)$  requires 
 vector gather loads,
which is memory inefficient.

Furthermore, the output subblock of matrix $B$
can be configured to be the same shape
for  the 2D5P star stencil,
i.e., $B_{n\times n}$
for both coefficient lines.
But for higher dimension star stencils,
$B$ has to be accessed
multiple times
with three or more orthogonal coefficient lines.
The reason is that the in-register subblock shape
must contain the direction of the 
coefficient line.
Take the 3D7P star stencil as an example,
for calculating  $CLS(*,1,1)$
the matrix shape must contain
the leftmost dimension
specified by $*$ in $CLS(*,1,1)$.
Therefore, it is legal to store 
either $B_{n\times 1\times n}$
or $B_{n\times n\times 1}$
in the matrix register
with the input vector
$A_{1\times 1\times n}$
or 
$A_{1\times n\times 1}$, respectively.
Then it is easy to see that 
we cannot complete the update with one matrix
shape for three or more coefficient line summations.

 In sum,  there is a trade-off
between the number of  outer product instructions
and the memory access pattern.
Specifically,
more orthogonal coefficient lines
results in more
inefficient data accesses of
input vectors of  $A$ and additional references of output subblocks of $B$, 
which may outweight the benefits of
less vector outer products.
Since the star
stencils can also be regarded as box stencils
as mentioned before, 
the non-zero weights can also be covered by parallel 
coefficient line.
We provide reasonable
coefficient line cover options 
and the corresponding theoretical analysis for star stencils.

\begin{figure}[t]
\centering
\centerline{\includegraphics[width=0.5\textwidth]{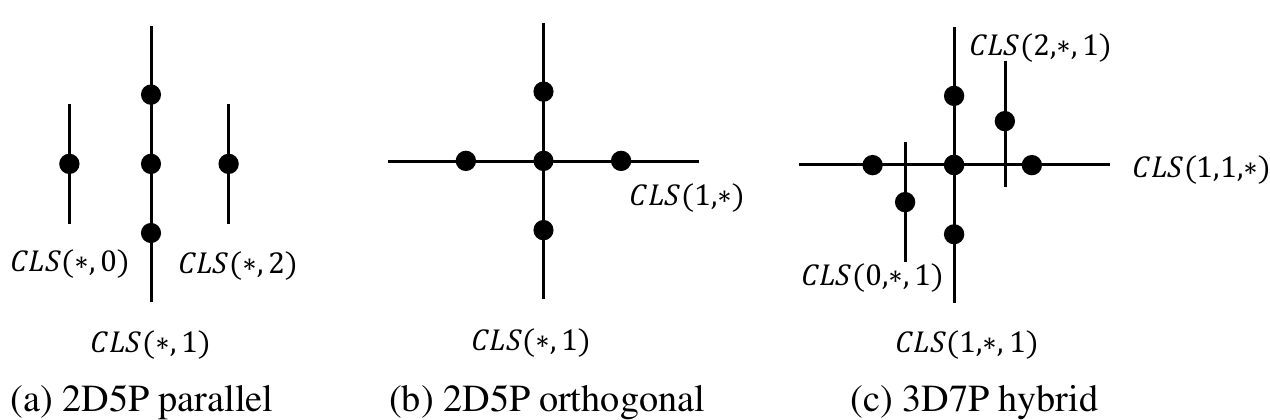}}
\caption{Coefficient line options for star stencils}
\label{fig-cls}
\end{figure}

Table \ref{cls-2d}
lists
two options, parallel and orthogonal,
for the 2D star stencil with order $r$.
The parallel option resembles the 
2D box stencil with order $r$
and includes $2r+1$ coefficient lines
$CLS(*,j)$ $(j=0,\dots,2r)$.
Figure \ref{fig-cls}(a)
illustrates for 
the 2D5P stencil ($r=1$).
The middle line $CLS(*,r)$ contains
$2r+1$ weights and consequently
generates $2r+n$ vector outer products.
All other coefficient 
lines have only one nonzero weight
and produce $n$ vector outer products.
Every input vector of $A$ consists of contiguous elements $A_{1\times n}$.
In the orthogonal option,
there are only two coefficient lines
and each one leads to $2r+n$ vector outer products.
Figure \ref{fig-cls}(b)
illustrates 
the 2D5P stencil.
The input vector for $CLS(r,*)$ has the form
$A_{n\times 1}$
that contains non-contiguous elements.

For  3D star stencils,
the parallel and orthogonal  
 options can be easily
obtained by extending those of
 2D star stencils.
As shown in Table \ref{cls-3d},
the numbers of
coefficient lines are
$4r+1$ and $3$ for the parallel
and orthogonal options, respectively.
All input vectors of $A$ in the parallel option
are contiguous accesses
and the output matrix can be set to the same shape
$B_{1\times n\times n}$.
In the orthogonal option,
$CLS(r,r,*)$ leads to strided accesses to $A$
and $CLS(*,r,r)$ demands $B_{n\times 1\times n}$,
which requires an extra
matrix reorganization.

\begin{table}[b]
\caption{Features of $CLS$ options for 3D star stencils}
\label{cls-3d}
\resizebox{\linewidth}{!}{
\begin{tabular}{cc|c|c|c}
\hline
\multicolumn{2}{c|}{Option}& Input & \#Outer Product   & Output \\ \hline
\multicolumn{1}{c|}{Parallel}& \begin{tabular}[c]{@{}c@{}}$CLS(i,*,k)$\\$i,k=0,...,2r$\\$ (i=r)\vee(k=r)$\end{tabular} & $A_{1\times 1\times n}$ & $(2r+n)+4r\cdot n$   & $B_{1\times n\times n}$ \\ \hline
\multicolumn{1}{c|}{\multirow{3}{*}{Orthogonal}} & $CLS(r,*,r)$  & $A_{1\times 1\times n}$ & \multirow{3}{*}{$3(2r+n)$} & $B_{1\times n\times n}$ \\
\multicolumn{1}{c|}{}& $CLS(r,r,*)$  & $A_{1\times n\times 1}$ &  & $B_{1\times n\times n}$ \\
\multicolumn{1}{c|}{}& $CLS(*,r,r)$  & $A_{1\times 1\times n}$ &  & $B_{n\times 1\times n}$ \\ \hline
\multicolumn{1}{c|}{\multirow{2}{*}{Hybrid}} &  \begin{tabular}[c]{@{}c@{}}$CLS(i,*,r)$ \\$i=0,...,2r$\end{tabular}   & $A_{1\times 1\times n}$ & \multirow{2}{*}{$2(2r+n)+2r\cdot n$} & $B_{1\times n\times n}$ \\
\multicolumn{1}{c|}{}& $CLS(r,r,*)$ & $A_{1\times n\times 1}$ &  & $B_{1\times n\times n}$ \\ \hline
\end{tabular}}
\end{table}

The non-contiguous input vector,
$A_{n\times 1}$ in 2D stencils
or $A_{1\times n\times 1}$
in 3D stencils,
can be loaded by a gather instruction,
% However, the type of memory access
which is inefficient.
Since we always ensure that a subblock of $B$ is
updated by both  non-contiguous vectors $A_{n\times 1}$ or $A_{1\times n\times 1}$
and contiguous vectors
$A_{1\times n}$ or $A_{1\times 1\times n}$,
it can employ matrix registers to 
produce these non-contiguous input vectors by a matrix transpose,
i.e., vector-to-matrix  moves in rows
and matrix-to-vector moves in columns.

The orthogonal option for 3D stencils
requires two output shapes,
$B_{1\times n\times n}$ and $B_{n\times 1\times n}$.
It has to perform 
redundant memory references
to $B$.
% either
% matrix transpose or gather load through memory accesses.
Thus the orthogonal option is expected to beat the parallel option
only for stencils with large orders.
To find a balanced solution, 
Figure \ref{fig-cls}(c) illustrates  a
hybrid option and 
the last row in Table \ref{cls-3d} lists its features.
This option completes the update with 
a single subblock shape $B_{1\times n\times n}$.
Its access pattern for the input vector
is similar to that of the orthogonal option,
and the output subblock to that of the parallel option.
The number of outer products is between
 those of the two options.

% $CLS(1,*,1)$,
% the input vector of $A$
% can hold contiguous elements
% i.e., $A_{1\times 1\times n}$.
% Therefore the output matrices
% are

% However, for $CLS(1,1,*)$,
%  $A_{1\times 1\times n}$.

% cover use $CLS(*,0)$, $CLS(*,1)$ and
% $CLS(*,2)$ to cover the coefficient matrix.

% As we prefer
% we try to 
% CLS(1,1,∗)

% According to Equation~(\ref{final}, 
% all the involved input vectors
% of $A$ hold contiguous elements.

% According to the aforementioned analysis,
% for a low order 2D star stencil,
% the benefit

% \begin{figure}[htbp]
% \centering
% \centerline{\includegraphics[width=0.5\textwidth]{data_access.pdf}}
% \caption{2d9p box}
% \label{fig-2}
% \end{figure}

% \begin{figure*}[b]
% \centering
% \centerline{\includegraphics[width=0.95\textwidth]{impletation.pdf}}
% \caption{Multi-dimensional unrolling and outer product scheduling}
% \label{fig-opt}
% \end{figure*}

\subsection{Multi-dimensional Unrolling}

\begin{figure*}[b]
\centering
\centerline{\includegraphics[width=0.95\textwidth]{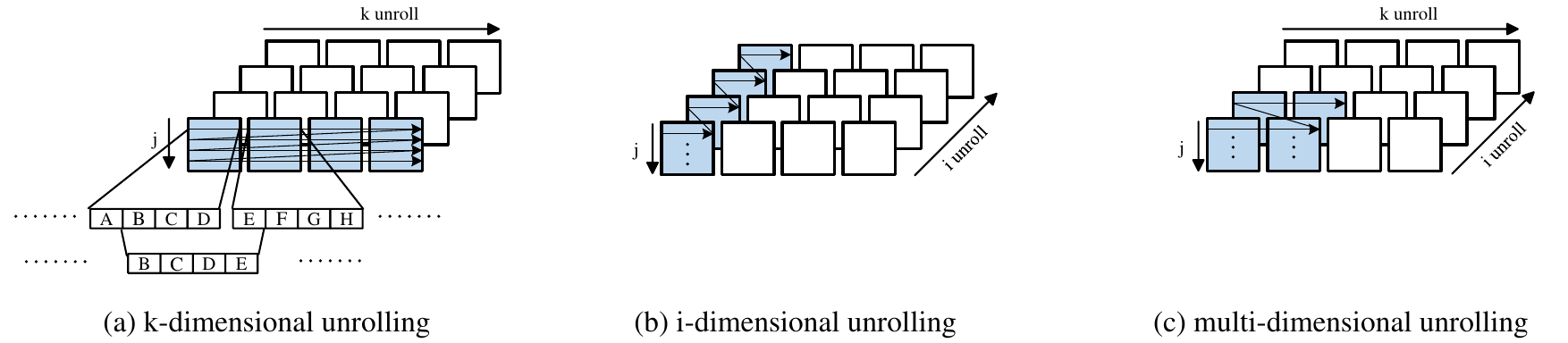}}
\caption{Multi-dimensional unrolling and outer product scheduling}
\label{fig-opt}
\end{figure*}

Loop unroll is a classic
optimization technique
to reduce the loop overhead and improve the memory 
accesses. 
 % enhance the data reuse.
For the proposed algorithm, 
the unrolled granularity
is the subblock of $B$.
As the contiguous memory access is
efficient than the non-contiguous one,
it  prefers to unroll
along the unit-strided dimension.
Figure \ref{fig-opt}(a)
illustrates the unrolling for 3D stencils with a unroll factor
$uk=4$.
The four contiguous blue subblocks along the $k$ dimension
are calculated and written to memory together
with a set of matrix registers.

Due to the low operational intensity, the stencil computation is often regarded as a 
memory-starving kernel.
Since neighbor elements lie 
% the stencil computation
% accesses  
along
all dimensions,
it is desirable
to explore more data reuses along
other directions.
Figure \ref{fig-opt}(b)
shows the unrolling 
along 
the $i$ dimension with a factor  $ui=4$.
An element may be scattered to all its neighbors and only needed to be loaded to the vector register only once.

According to the formula, an input vector is scattered to neighbor rows of the matrix register by a single outer product.
% filled with a subblock of output $B$
Therefore it
actually indicates an implicit unroll, 
i.e., an input data may be reused along the $i$ dimension in 2D stencils or the $j$ dimension in
3D stencils. 
This dimension is not further unrolled,
as we do not observe a performance improvement with it
for both 2D and 3D stencils.

Generally, loop unroll is 
only applied to the innermost loop
or the unit-strided dimension.
With its dependence pattern, 
the stencil computation
desires data reuses over
all dimensions.
For example,
the tiling techniques often
block the stencil computations along multiple space dimensions at the cache level.
However, we are not aware of any multiple unrolling methods for stencil computations.
The reason is that the register resources are limited compared with
the large cache size. 
Fortunately, with additional  matrix registers capable of holding
2-dimensional subblocks,
multi-dimensional unrolling
is feasible.
Figure \ref{fig-opt}(c)
pictures a two-dimensional unrolling
over the $i$ and $k$ dimensions
with factors $ui=uk=2$.
% The optimal parameters
% are determined experimentally and will be discussed in the evaluation section.

% i.e., the loop body
% performs $u$ times more instructions
% along the innermost loop index
% with an unroll factor $u$.

\subsection{Outer Product Scheduling}

% Furthermore, this granularity 
% is coarse.
% According to the final formula,
% the innermost loop body
% is a block of the output matrix.

With multi-dimensional unrolling,
there exists  $ui\times uk$ subblocks of $B$ in matrix registers.
The naive scheme that updates them 
one after another is
inefficient.
Take the 2D9P box stencil as an example,
all the 3 coefficient lines require
total $3(n+2)$ coefficient vectors.
It fails to reuse them across  
subblocks 
and has to reload them for each subblock.

Furthermore, 
each outer product  actually scatters an input vector
to multiple rows of a subblock of $B$, i.e.,
to all its neighbors along the  $j$ dimension
for 3D box stencils.
Similarly,  one input vector may also be reused by adjacent $ui$ subblocks 
of $B$ along the $i$ dimension.
The naive scheme
also fails to utilize this type of reuse and incurs multiple loads of the same input vector.

To address these problems, 
we reschedule the outer products of all in-register subblocks.
Since a coefficient vector corresponds
to input elements from a plane $A_{*,j,*}$, we resort to 
load all needed input vectors within a plane and  scatter them 
to all matrix registers.
Therefore, the coefficient vector storage can be reused to fill the
next ones for updating with  $A_{*,j+1,*}$ in the next plane.
For the $i$ dimensional reuse of input vectors,
It is  desirable to 
continuously update all other  $ui-1$ neighbor $B$ subblocks after
obtain an input vector.

In sum, our scheduling groups the vector products according to the input vectors 
of $A$, i.e., sort according to 
firstly scatter one input vector
along the $i$ dimension,
assemble input vector along the $k$
dimension and maximal reuse registers for coefficient vectors across $j$ planes.
The arrows in the
unrolled blue subblocks in Figure~\ref{fig-opt}
illustrate this scheduling.

The final problem is the scheduling along the unit-strided dimension.
One well-known problem induced by the vectorization of stencils is the 
\emph{data alignment conflict}. 
It arises from the fact that continuous vectors require that the same value appears at different positions of vectors. 
In this work, we adopt the 
data reorganization method.
All the needed elements of $A$
are loaded to vector registers
and all the input vectors
involved in outer products
are derived through inter-register
assembling.
Figure~\ref{fig-opt}(a)
presents an example where
the two loaded vectors
$(A,B,C,D)$ and $(E,F,G,H)$
are combined to obtain the needed input vector $(B,C,D,E)$.

\begin{algorithm}[b] %如果不能显示，这里要把[H]给加上就行了“H”是指定伪代码浮动体的位置
  \caption{Updating the first group of unrolled subblocks for  3D box stencils with order $r$ }
  \label{alg}
  \For{$i\in 0,\dots,ui-1$}
  {
\For{$k\in 0,\dots,uk-1$}
{
$BM_{i,k} \leftarrow 0 $
}
  }  
  \For{$j\in -r,\dots,n+r-1$}
  {
\For{$io\in -r,\dots,r$}
{
\For{$ko\in -r,\dots,r$}
{
assemble  $CV_{io,ko}$
}
}
\For{$i\in -r,\dots,ui+r-1$}
{
load $A_{i,j,-r:(uk\cdot n+r-1)}$ 

\For{$ko\in -r,\dots,r$}
{
\For{$k\in 0,\dots,uk-1$}
{{\small
assemble $AV=A_{i,j,(k\cdot n+ko):(k\cdot n+ko+n-1)}$}

\For{$io\in -r,...,r$}
{
\If{$0\leqslant i+io< ui$}
{
 $BM_{(i+io),k} \xleftarrow{accum} CV_{io,ko}\otimes AV$
}
}
}
}
}
  }  
  \For{$i\in 0,\dots,ui-1$}
  {
\For{$k\in 0,\dots,uk-1$} 
{
store $BM_{i,k}$
}
  }

\end{algorithm}

\subsection{Put It All Together}

Algorithm \ref{alg}
shows the pseudo-code 
for the 3D box stencil with order $r$.
The first nested loop in Line 1 to Line 3 sets
the multi-dimensional unrolled $ui\times uk$ matrix registers $BM$
  to zero.
Without loss generality,
we assume the updated elements are
$B_{0:(ui-1),0:(n-1),0:(uk*n-1)}$,
i.e., the first group of  $ui\times uk$ subblocks of $B$.

Following the scheduling scheme discussed above, 
the middle nested  loop in Line 4 to Line 15
 assembles  the coefficient vectors
$CV$ and input  vectors $AV$, 
and performs the outer products.
The outermost loop in Line 4
iterates over all related $j$ indices, i.e., all $j$ planes.
Inside each iteration, it firstly assembles $(2r+1)^2$ coefficient vector $CV$
for the current $j$
in Line 5 to Line 7.
We omit the details, which are straightforward
with the final formulas in Equation~(\ref{co}) and 
(\ref{final}).
All the following outer products
in this iteration only use
these $CV$s to product with the input elements in
the plane.

The inner loop in Line 8
iterates over all indices in the $i$ dimension
and each iteration loads all needed elements
$A_{i,j,-r:(uk\cdot n+r-1)}$  to vectors.
The next double-nested loop in Line 10 and 11
assembles every involved input vector $AV$
by the data reorganization of  vectors
loaded in Line 12.
Each $AV$ is scattered to all
$BM$s in the innermost loop in Line 13
if the $i$ index falls into the range of
$0$ to $ui-1$.
Finally, the last nested loop in Line 16 to 18 writes
all  subblocks $BM$ to the output $B$.

Algorithms for other stencil types and dimensions
are similar. 
We also implemented a code generator using Python.
It accepts the stencil type, the coefficient line option,
and  unroll factors,
and automatically creates
the final codes.
Note that the generated code
only keeps the loops in Line 4 and Line 8,
and fully unrolls all other loops.
Therefore the branch in Line 14 is eliminated.

\section{Evaluation}

\subsection{Methodology}

We conducted the experiments on a proprietary
ARM simulator, whose key 
parameters are configurable.
The number of  outer production
unit is set to 1.
% This is consistent to the 
% common configurations
% of existing processors,
% e.g., the Intel
% Sapphire Rapids CPU
%  deploys  one TMMU per core.
 The vector length is set to 512-bit,
holding %16 single-precision or
8 double-precision floating-point numbers. 
The size of the matrix register
is $8\times 8$.
The register number for vector
is 32 and 8 for matrix.
% the $SVL_B$ is also 512-bit length
% and therefore the matrix register size is  
%$16\times 16$  for single-precision or 
% for double-precision floating-point numbers.
% The total number of matrix register is
% 8 for double-precision %or 4 for single-precision 
% floating-point numbers.
Resemble to Kunpeng 920 CPU,
the memory hierarchy in CPU consists of a 64KB  L1 data cache and 64KB  instruction cache, and a 
 512 KB private L2 cache.
 We tested box and star stencils in 2D and 3D with various orders in double-precision type.
 The  compiler is Bisheng 2.1.0 which is based on the clang version 12.0.0.

% and
% its  frequency
% is 80\% of that of t

% AVX-512 is reduced frequncy
% compared with AVX2

 \subsection{Results}

\begin{figure*}[t]
  \centering
 \subfloat[2D, size=$64^2$] {
   \includegraphics[width=0.24\linewidth]{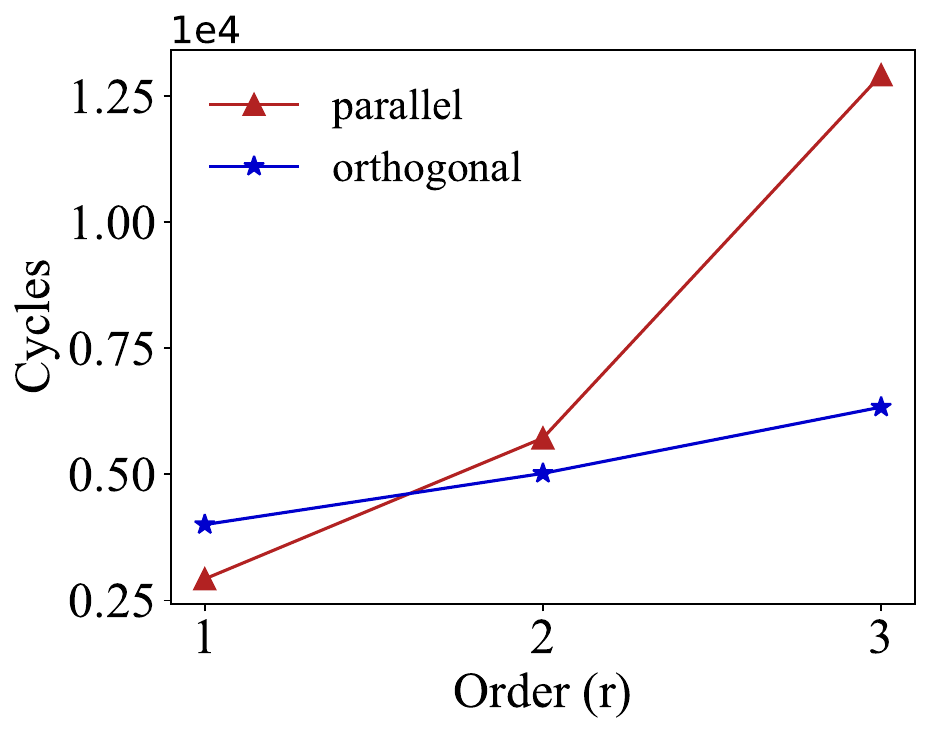}
   \label{result-opt1-1}
  } 
\hfill
% \subfloat[2D $128^2$] {
%    \includegraphics[width=0.225\linewidth]{cls-2d-128.pdf}
%    \label{fig-2d-9p-jacobi-2}
%   }
%   \hfill
% \subfloat[2D $256^2$] {
%    \includegraphics[width=0.235\linewidth]{cls-2d-256.pdf}
%    \label{fig-2d-life-1}
%   } 
% \hfill
\subfloat[2D, size=$512^2$] {
   \includegraphics[width=0.235\linewidth]{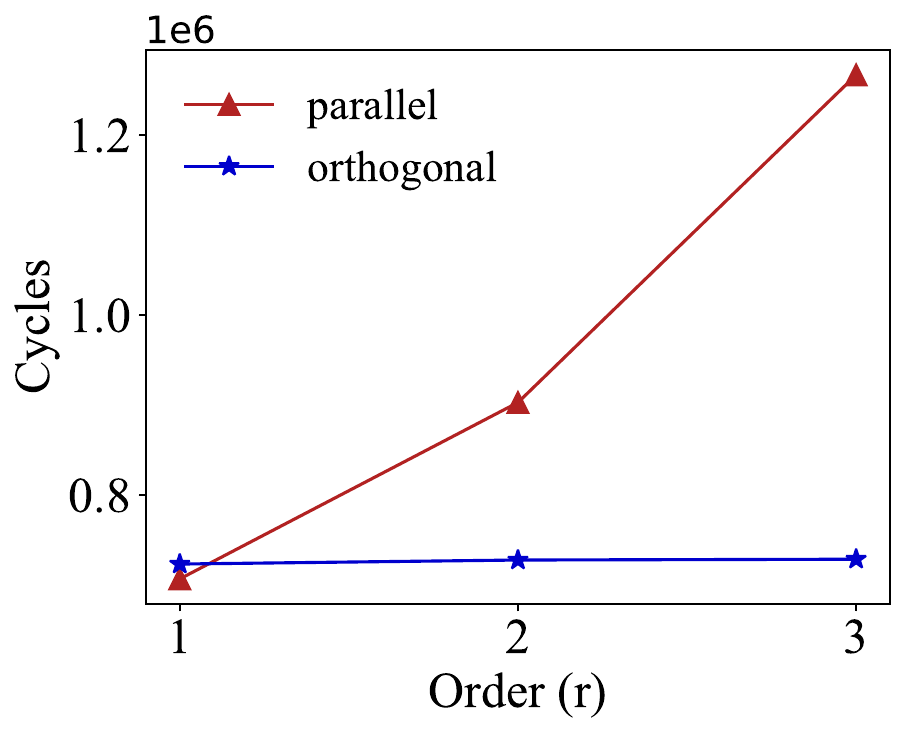}
   \label{result-opt1-2}
  }
\hfill
 \subfloat[3D, size=$8^3$] {
   \includegraphics[width=0.23\linewidth]{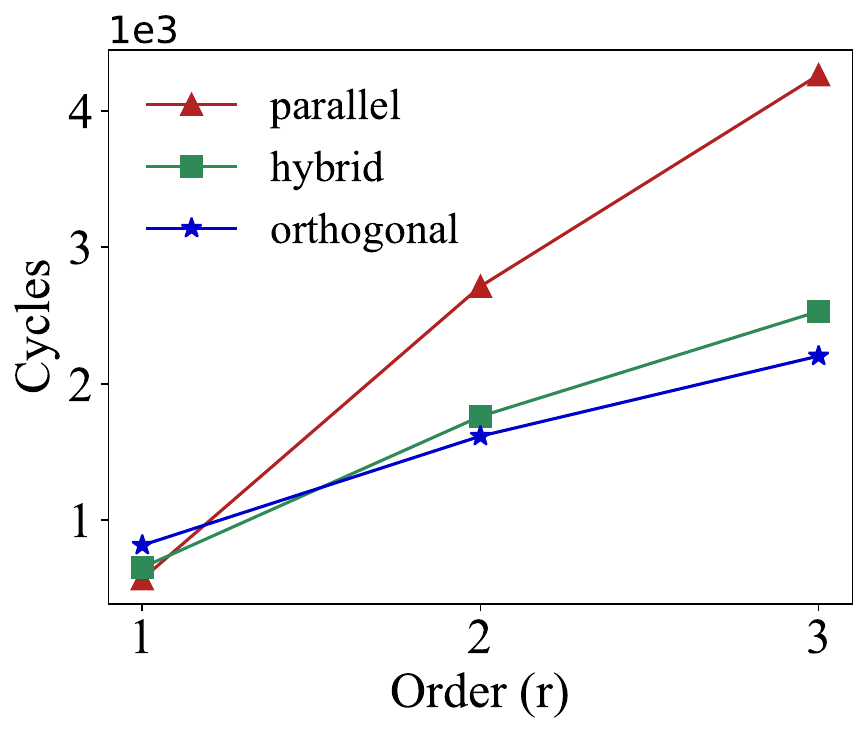}
   \label{result-opt1-3}
  } 
\hfill
% \subfloat[3D $16^3$] {
%    \includegraphics[width=0.23\linewidth]{cls-3d-16.pdf}
%    \label{fig-2d-9p-jacobi-2}
%   }
%   \hfill
% \subfloat[3D $32^3$] {
%    \includegraphics[width=0.23\linewidth]{cls-3d-32.pdf}
%    \label{fig-2d-life-1}
%   } 
% \hfill
\subfloat[3D, size=$64^3$] {
   \includegraphics[width=0.23\linewidth]{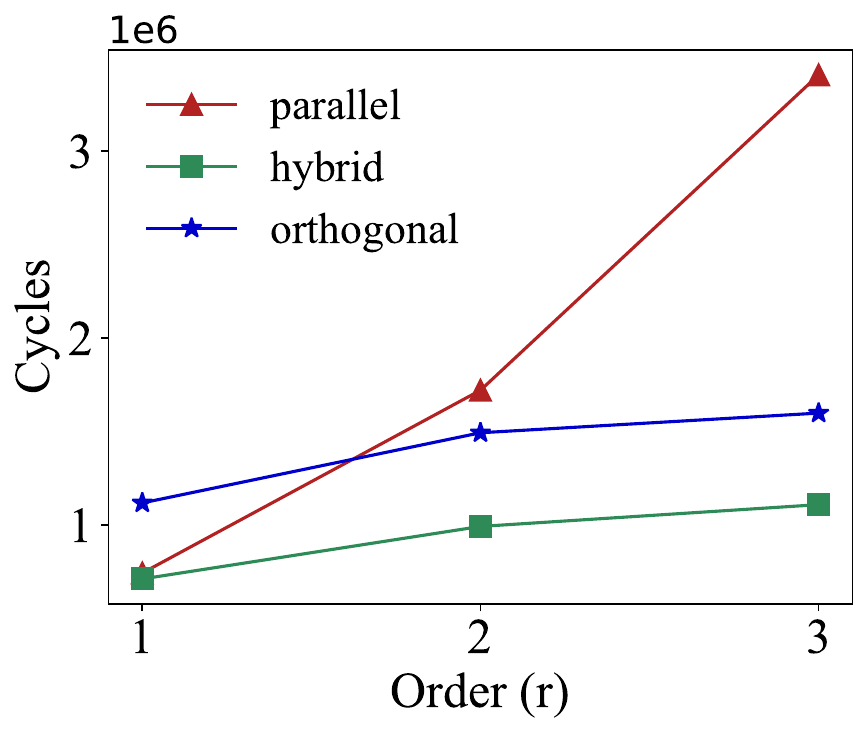}
   \label{result-opt1-4}
  }
  \caption{Performance of star stencils with various coefficient line options}
\label{result-opt1}
\end{figure*}

\begin{figure*}[t]
  \centering
  % \centerline{\includegraphics[width=0.4\textwidth]{legend.png}}
 \subfloat[2D, size=$64^2$] {
   \includegraphics[width=0.235\linewidth]{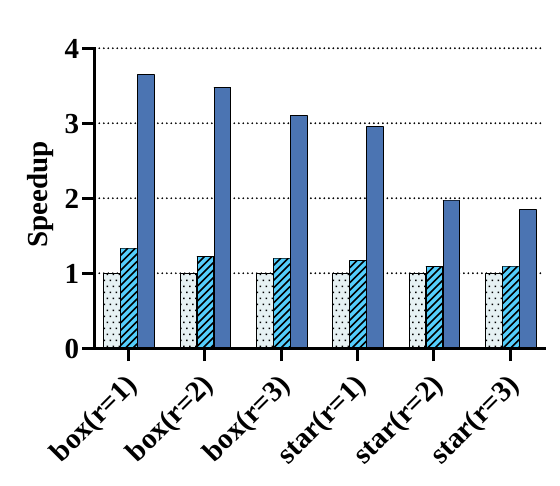}
   \label{result-opt2-1}
  } 
\hfill
\subfloat[2D, size=$512^2$] {
   \includegraphics[width=0.235\linewidth]{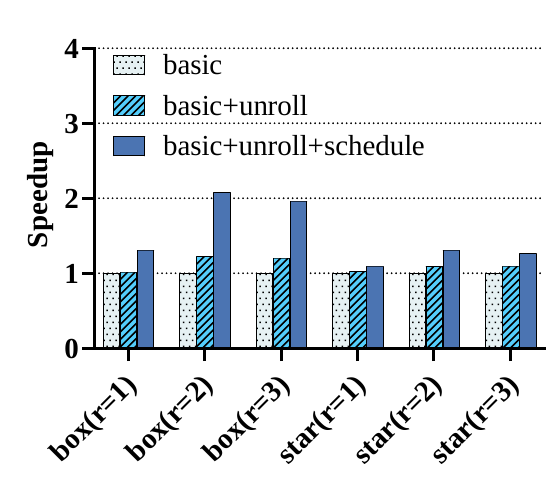}
   \label{result-opt2-2}
  }
  \hfill
\subfloat[3D, size=$8^3$] {
   \includegraphics[width=0.235\linewidth]{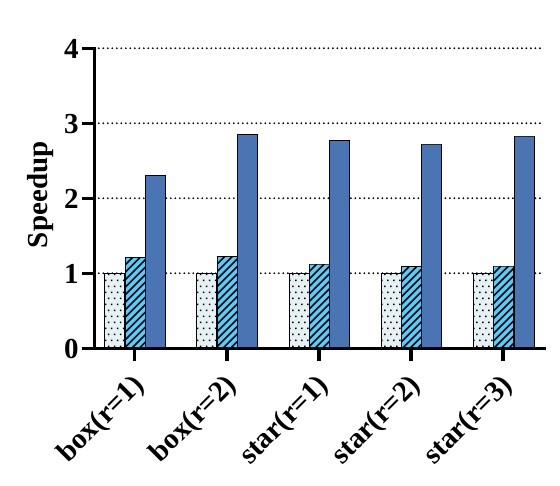}
   \label{result-opt2-3}
  } 
\hfill
\subfloat[3D, size=$64^3$] {
   \includegraphics[width=0.235\linewidth]{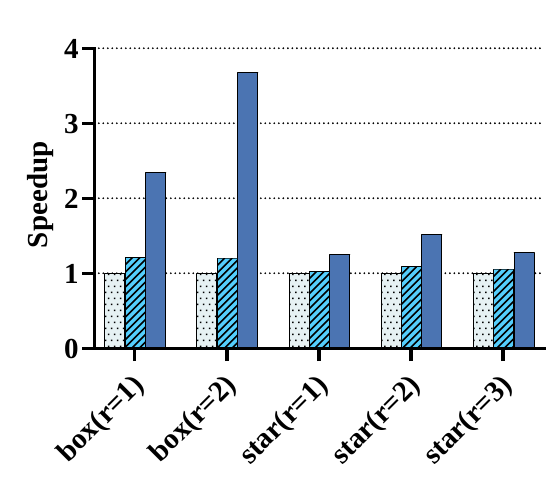}
   \label{result-opt2-4}
  }
  \caption{Improvement of multi-dimensional unrolling and outer product scheduling}
\label{result-opt2}
\end{figure*} 
% 图3表示了不同star stencil的cls选项随着order的不同的性能的改变。

\begin{figure*}[t]
  \centering
 \subfloat[2D9P box] {
   \includegraphics[width=0.235\linewidth]{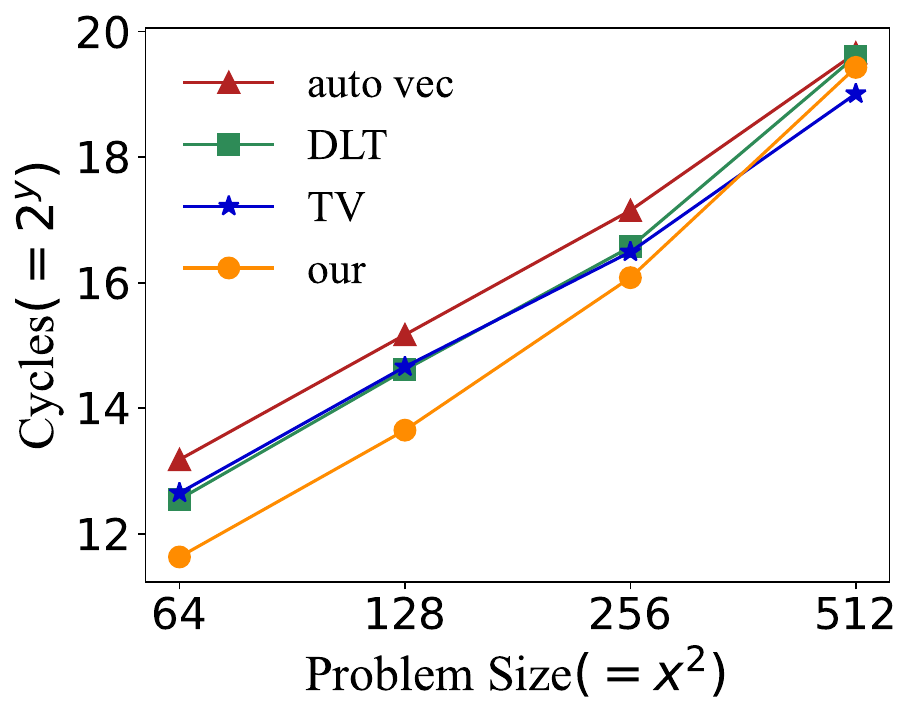}
   \label{result-compare-1}
  } 
\hfill
\subfloat[2D5P star] {
   \includegraphics[width=0.235\linewidth]{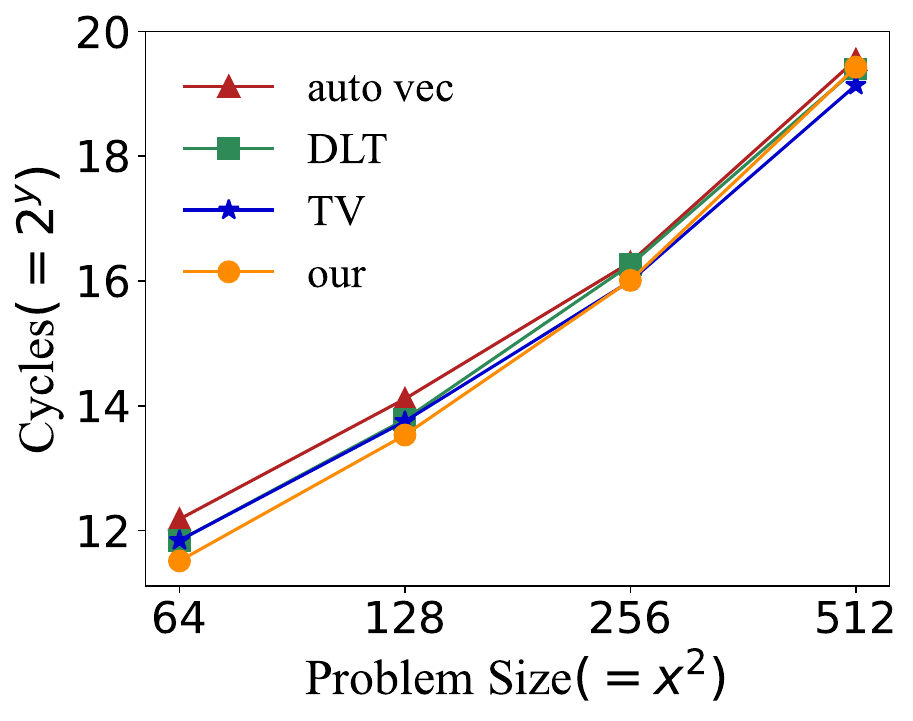}
   \label{result-compare-2}
  }
  \hfill
\subfloat[3D27P box] {
   \includegraphics[width=0.235\linewidth]{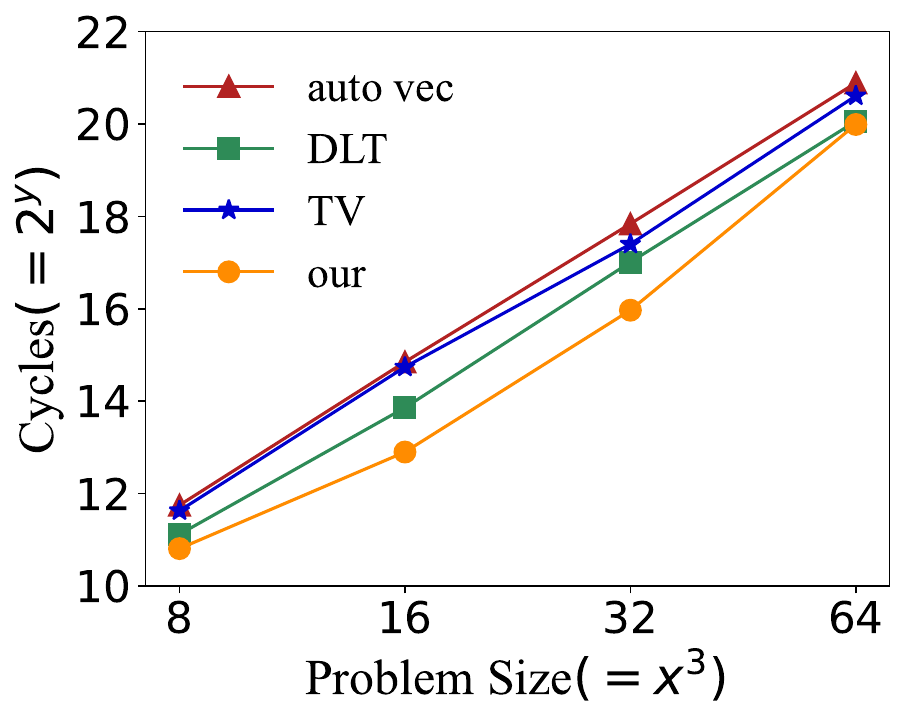}
   \label{result-compare-3}
  } 
\hfill
\subfloat[3D7P star] {
   \includegraphics[width=0.235\linewidth]{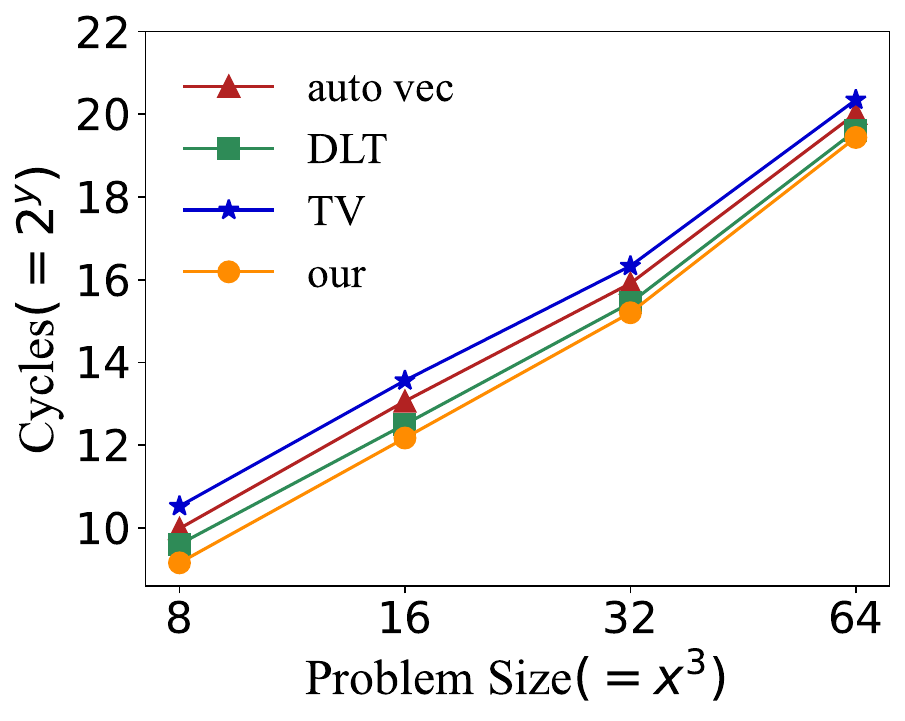}
   \label{result-compare-4}
  }
  \caption{Performance comparison with existing vectorization methods for stencils with $r=1$}
\label{result-compare}
\end{figure*} 

\begin{table*}[t]  
\begin{center}
\caption{Speedup comparison (normalized to auto-vectorization, best in grey)}
\label{tab-result1}
\setlength{\tabcolsep}{1.5mm}{
\begin{tabular}{c|cccc|cccc|cccc|cccc}
\hline
 \multirow{2}{*}{2D}&  \multicolumn{4}{c|}{$64^2$} &  \multicolumn{4}{c|}{$128^2$}  & \multicolumn{4}{c|}{$256^2$} & \multicolumn{4}{c}{$512^2$} \\
 \cline{2-17}
 & DLT & TV & \multicolumn{2}{c|}{our (option)} 
 & DLT & TV & \multicolumn{2}{c|}{our (option)} 
 & DLT & TV &\multicolumn{2}{c|}{our (option)}
 & DLT & TV &\multicolumn{2}{c}{our (option)}\\
 \hline
 box ($r=1$)   & 1.55 & 1.44 & \cellcolor{mygray}2.92  & ($j8$) & 1.48 & 1.43 & \cellcolor{mygray}2.87 & ($j8$) & 1.50 & 1.58 & \cellcolor{mygray}2.11  & ($j8$) & 1.03 & \cellcolor{mygray}1.56 & 1.17 & ($j8$)\\
box ($r=2$)   &1.48 & 2.43& \cellcolor{mygray}4.58  & ($j8$) & 1.43 & 2.09 & \cellcolor{mygray}3.78 & ($j8$) & 1.42 & 2.13 & \cellcolor{mygray}3.91  & ($j8$) & 1.02 & 1.80 & \cellcolor{mygray}2.17 & ($j8$)\\
box ($r=3$)   &1.51 & 2.47 & \cellcolor{mygray}4.71  & ($j8$) &1.33 & 1.67 & \cellcolor{mygray}3.04 & ($j8$) & 1.39 & 2.05 & \cellcolor{mygray}4.33  & ($j8$) & 1.09 & 1.79 & \cellcolor{mygray}2.26 &($j8$)\\
star ($r=1$)   & 1.27 & 1.27 & \cellcolor{mygray}1.59  & ($p$-$j8$) & 1.26 & 1.28 & \cellcolor{mygray}1.50 & ($p$-$j8$) & 1.02 & \cellcolor{mygray}1.23 &  1.22  & ($p$-$j8$) & 1.11 & \cellcolor{mygray}1.33 & 1.09 &($p$-$j8$)\\
star ($r=2$)   & 1.28 & 1.31 & \cellcolor{mygray}1.48  & ($o$-$j4$) & 1.29 & 1.32 & \cellcolor{mygray}1.43 & ($o$-$j4$) & 1.30 & 1.39 & \cellcolor{mygray}1.41  & ($o$-$j4$) & 1.03 & \cellcolor{mygray}1.33 & 1.19 & ($o$-$j4$)\\
star ($r=3$)   & 1.30 & 1.29 & \cellcolor{mygray}1.64  & ($o$-$j4$) & 1.32 & 1.34 & \cellcolor{mygray}1.70 & ($o$-$j4$) &  1.29 & 1.42 & \cellcolor{mygray}1.65  & ($o$-$j4$) & 1.04 & \cellcolor{mygray}1.42 & 1.02 & ($o$-$j4$)\\
\hline
 3D &  \multicolumn{4}{c|}{$8^3$} &  \multicolumn{4}{c|}{$16^3$}  & \multicolumn{4}{c|}{$32^3$} & \multicolumn{4}{c}{$64^3$} \\
\hline
box ($r=1$)   & 1.55 & 1.09 & \cellcolor{mygray}1.93 & ($i4$)  & 1.98 & 1.08 & \cellcolor{mygray}3.85& ($i4k2$) &  1.78 & 1.36 & 3\cellcolor{mygray}.68& ($i4k2$)  & 1.77 & 1.21 & \cellcolor{mygray}1.87 & ($i4k2$) \\
box ($r=2$)  &1.03 & 1.13 & \cellcolor{mygray}3.90 & ($i4$) & 0.86 & 1.21 &  \cellcolor{mygray}3.44 & ($i4k2$) & 1.01 & 1.19 & \cellcolor{mygray}3.88& ($i4k2$) &1.54 & 1.17 &  \cellcolor{mygray}3.05 & ($i4k2$)\\
star ($r=1$) &1.30 & 0.69 & \cellcolor{mygray}1.77& ($p$-$i8$) & 1.47 & 0.71 & \cellcolor{mygray}1.64 & ($p$-$i8$) & 1.39 & 0.75 & \cellcolor{mygray}1.46& ($p$-$i4k2$)  & 1.33 & 0.80 & \cellcolor{mygray}1.50 & ($h$-$k4$) \\
star ($r=2$)   & 2.18 & 1.37 & \cellcolor{mygray}3.73& ($o$-$i4$)  & 2.89 & 1.43 & \cellcolor{mygray}3.37& ($o$-$i4$) & 2.76 & 1.32 &  \cellcolor{mygray}3.10& ($o$-$i4$)  & 1.17 & 1.01 & \cellcolor{mygray}1.81 & ($h$-$k4$) \\
star ($r=3$)   & 3.13 & 1.51 & \cellcolor{mygray}4.14& ($o$-$i4$)  & 2.61 & 1.46 & \cellcolor{mygray}3.63 & ($o$-$i4$) & 2.45 & 1.35 & \cellcolor{mygray}3.04 & ($o$-$i4$) & 1.38 & 1.05 & \cellcolor{mygray}1.80 & ($h$-$k4$) \\
\hline
\end{tabular}}
\end{center}
\end{table*} 
Figure~\ref{result-opt1} shows the performance of star stencils with different coefficient line options.
The parallel option obtains the best performance for order=1 in all cases.
However, 
the orthogonal and hybrid curves are flatter than the parallel curve as the order increases.
This is due to the higher growth rate of the outer product number for the parallel option. 
According to Table \ref{cls-2d} and \ref{cls-3d},
the increases are $O(n)$ and $O(1)$ for the parallel and orthogonal options, respectively.
Similarly, the parallel option also incurs
more data transfers than the orthogonal one as the order increases.

Figure~\ref{result-opt1-1} and \ref{result-opt1-2}
exhibit the performance of the in-cache problem size of $64^2$ and the out-of-cache size of $512^2$ for 2D star stencils, respectively.
In both cases, the orthogonal option is faster
for orders larger than 1.
For stencils with the out-of-cache problem size, 
the orthogonal curve is even flatter
than that with in-cache size.
The reason is that it has little influences
on the memory performance
as the order increases.
Specifically, with the order increased by 1, each
subblock of $B$ requires two more boundary rows.
Since the L1 cache can hold 16 rows 
of 512 double-precision floating-point numbers,
it does not raise the memory transfer volume.
Thus we expect little performance change.

% Figure~\ref{result-opt1-3} and \ref{result-opt1-4}
% show the in-cache and out-of-cache performance 
For 3D star stencils,
% For high orders the parallel 
% option is the fastest for the small problem sizes.
the orthogonal option in Figure~\ref{result-opt1-4}
is not that flatter as the 2D case in Figure~\ref{result-opt1-2}.
The reason is that the L1 cache can hold two $64^2$ planes and additional boundary planes with a higher order require
more L2 cache or memory accesses.
For large problem size, the hybrid option beats the
other two
for high orders.
The reason is similar since the latter leads to additional data reorganizations to subblocks of $B$.
This extra overhead is significant for out-of-cache data.

Figure ~\ref{result-opt2} shows the effectiveness of the multi-unrolling and instruction scheduling optimizations. 
Although the unrolling seems to have limited effects
in all cases, the instruction scheduling is
actually built upon it.
All star stencils adopt the best coefficient line options indicated in Figure~\ref{result-opt1}.
For 2D stencils, the unrolling only applies over the
contiguous $j$ dimension.
The unroll factors $uj$ are $8$ and $4$ for parallel options (all box stencils and star stencils with order 1) and 
orthogonal options (star stencils with high orders), respectively.
These options are also listed in brackets in Table~\ref{tab-result1}.
$p$-$j8$ means the parallel coefficient line and $uj=8$.
$o$-$j4$ denotes the orthogonal coefficient line and $uj=4$.
Note that all box stencils employ the parallel option and thus are omitted.

For the small problem size fit in the cache,
Figure~\ref{result-opt2-1}
and \ref{result-opt2-3} show  substantial improvements
for all stencils.
Since the instruction scheduling optimizes the data access and data reuse pattern
and apparently box stencils generate more input vectors,
star stencils achieve fewer speedups compared with box stencils.
For the large problem size,
Figure~\ref{result-opt2-2}
and \ref{result-opt2-4}
exhibit similar trends.
A performance model is desired to determine the optimal option and make a detailed analysis, which is left as future work.

% First, for 2D stencil, schedule bring a larger promotion for box shape stencils than star shape stencils because there is a data rearrangement to reduce data load in box shape stencil. 2D5P stencil also has a large promotion because it is calculate as parallel, which means it also has a data rearrangement. And the promotion of schedule is more obvious when problem size are $64^2$ than $512^2$ because when problem size is $512^2$ memory boundary turns to be the bottleneck.

Figure  \ref{result-compare} shows
the performance of the compiler's automatic vectorization,
DLT \cite{Henretty+:cc11},
temporal vectorization (TV) \cite{yuan2021temporal} and our method
on 2D and 3D stencils with order 1.
It also includes two more problem sizes that fit in cache, $128^2$ and $256^2$ for 2D,
and $16^3$ and $32^3$ for 3D.
Since box stencils incur more data reuse
than star stencils,
our method achieves more obvious improvements
for them
on problem sizes fit in cache.
Table \ref{tab-result1} lists
speedups over the auto-vectorization on all stencils.
Our method achieves the best performance on all 3D stencils.
The TV method extremely reduces the memory volume up to a
fourth and it obtains better improvement for 
the out-of-cache problem size on some 2D stencils.
The DLT speedups are consistent with the values in \cite{Henretty+:cc11}.
The TV has limited or negative speedups for 3D stencils,
as presented in~\cite{yuan2021temporal}.

Overall, the speedups
of all methods increase as the order increases for all box and star stencils in 2D and 3D.
For 2D and 3D box stencils, the speedups are
in the range of 3 to 5.
The 2D star stencils achieve smaller speedups around $1.5 \times$.
On the contrary, the 3D star stencils get better improvements due to the $i$-dimensional reuse.
% 3D star stencils attain similar improvements to 3D box stencils, while the with 2D star stencils to 2D box stencils. The reason is that the hybrid option balance
% the computation cost and memory reference pattern.
As for the optimization,
all star stencils adopt the same coefficient line options as those in Figure~\ref{result-opt1-1}
and \ref{result-opt1-3}
 for the middle two problem sizes.
For 2D stencils, the unroll factors
are also the same values.
All 3D stencils adopt the $i$-dimensional
unroll
except for the 3D star stencils on the largest problem size, which accepts the hybrid
coefficient line option.
For our method, all stencils obtain the maximal speedups
with the smallest problem size, except for the 3D27P box stencil that gets a speedup of $1.93\times$.
The reason may be the small total space size
and for $16^3$ it delivers a better improvement of $3.85\times$. 
For larger problem sizes beyond the values in Table
\ref{tab-result1},
we can expect similar speedups for our method
by utilizing a blocking scheme with similar blocking sizes.

\section{Related work}

To improve performance, 
vectorization and tiling
 are the two most powerful
transformation approaches.
Universal  vectorization techniques 
\cite{Allen.Kennedy:toplas87,Allen.Kennedy:book01,Sreraman.Govindarajan:ijpp00,Larsen.Amarasinghe:pldi10,Hampton.Asanovic:cgo08,Nuzman.Zaks:pact08,Nuzman.Zaks:pact08,Yount:hpcc15,zhao2019exploiting}
are  extensively studied
by the compiler community.
Sophisticated methods~\cite{Henretty+:cc11,Caballero+:ics15}
targeting stencil computations
are also proposed.
This work utilizes the
data reorganization technique
\cite{Zhou.Xue:cgo16} to reduce the 
memory references of the input array.

Tiling \cite{Irigoin.Triolet:popl88,McKeller.Coffman:cacm69,Lam+:asplos91,Wolf.Lam:pldi91,Wolfe:sc89,Wonnacott.Strout:impact13}
aims
to explore the data locality and parallelism of multiple loop nests.
There exists an extremely large number of tiling
methods for the stencil computation,
e.g., overlapped tiling \cite{Ding.He:sc01,Rastello.Dauxois:ipdps02,Rivera.Tseng:sc00,Nguyen+:sc10,Meng.Skadron:ics09,Krishnamoorthy+:pldi07,Holewinski+:ics12,Philips.Fatica:ipdps10},
time skewing \cite{Song.Li:pldi99,Wonnacott:ijpp02,Jin+:sc01,Andonov+:spaa01,Andonov+:tpds03},
diamond tiling \cite{Bondhugula+:pldi08,Bandishti+:sc12,Pananilath+:taco15,Grosser+:ppl14},
cache-oblivious tiling
\cite{Frigo.Strumpen:ics05,Frigo.Strumpen:spaa06,Strzodka+:ics10,Tang+:spaa11},
split tiling
\cite{Wonnacott:ipdps00,Krishnamoorthy+:pldi07,Cui+:jcst10,Zhou+:cgo12,Grosser+:gpgpu13,Henretty+:ics13} and
hybrid tiling \cite{Strzodka+:icpp11,Malas+:siamjosc15,Grosser+:cgo14}.
% Wonnacott and Strout present a comparison on the scalability of many existing tiling schemes \cite{Wonnacott.Strout:impact13}.
This work is orthogonal to tiling techniques.
% , this work focuses on map the stencil computation in one time step on the vector outer product unit.
According to the experiment results, it is desirable to reuse data blocks 
over several time steps and achieve
a higher speedup with the in-cache data.
A combination of the two techniques is our future work.

% the multiple load method is often used by
%  by production compilers. 
%  It loads all the needed vectors from memory straightforwardly.
% It incurs the well-known data alignment conflict.
% Another milestone approach to address the data alignment conflict for stencil computations is the Dimension-Lifting Transpose method \cite{Henretty+:cc11}. 
% It turns to put the points with read-read dependencies in the same position of different vectors.
% However, it's hard to implement the  transpose in-place and it often chooses to use 
% an additional array to store the transposed data. 
% The data reorganization vectorization \cite{Caballero+:ics15,Zhou.Xue:cgo16} 
%  loads each input element to vector register only once and assembles the required vectors 
% via inter-register data permutations instructions. Compared with the multiple load method, 
% this data permutations method reduces the memory bandwidth usage and takes the advantage of the rich set of data-reordering instructions supported by most SIMD architectures. 

Several work \cite{Moreira+:arxiv21,Liu+:ics22}  has studied the 
transformation of the stencil computation 
to matrix computations.
TCstencil \cite{Liu+:ics22} is the seminal work that
adapts the stencil computation to matrix-matrix multiplications.
However, their method is developed from the 
gather view of the stencil definition and only applies to two-dimensional stencils.
Since TCstencil targets the Tensor cores
on GPU, it is not amenable
to the instruction scheduling optimization
that reduce both the coefficient and input vector
reorganizations.
Furthermore, TCstencil 
does not map all the stencil computation on matrix units as we do and
it employs the traditional vector arithmetic instructions
to perform some calculations.
Moreira et al. \cite{Moreira+:arxiv21}
studied the convolution, a similar computation pattern to the stencil
on IBM's Matrix-Multiply Assist
with the outer product instructions.
However, it is also derived from
the gather mode
and no experimental result is reported.

\section{Conclusion}
We have presented a new algorithm for stencil computation with outer products. It arises from the scatter view and
centers on a key concept of
coefficient line.
A set of optimizations
is further proposed 
to improve the memory reference pattern, execution pipeline and data reuse.
Evaluation on a  simulator shows that it achieves a maximal speedup of $4.71\times$ compared with the vectorized algorithm.

\bibliography{stencil2}

% %% Appendix
% \appendix
% \section{Appendix}

% Text of appendix \ldots

\end{document}